\documentstyle[12pt,epsf]{article}
\title{QCD Corrections to  $b \to s e^+e^-$ Decay } 
\author{ Cai-Dian L\"{u}$^a$ and Han-Wen Huang$^b$\\
a Department of Physics, Technion-Israel Institute of Technology,\\
Haifa 32000, Israel;\\
b Institute of Theoretical Physics, Academia Sinica, P.O.Box 2735,\\
 Beijing 100080, China.}

\date{}

\begin{document}
\maketitle
\begin{picture}(0,0)(0,0)
\put(320,290){{\large hep-ph/9608235}}
\put(320,270){{\large TECHNION-PH-96-18}}
\put(320,250){{\large August, 1996}}
\end{picture}

\begin{abstract}
We give a more complete calculation of $b \to s e^+e^- $ decay
including leading log QCD corrections from $m_{top}$
to $M_W$ in addition to corrections  from $M_{W}$ to $m_b$.  
The differential decay rate is found to be slightly suppressed for a large
invariant mass of the $e^+e^- $ pair; while the integrated width is slightly
enhanced comparing with the results without the 
QCD running from  $m_{top}$ to $M_W$. 
\end{abstract}


\newpage

		\section{Introduction}

The rare decay $b \to s e^+e^-$ is one of the very useful channels for the 
study of models beyond standard model. This deeply depends on more precise 
calculations of this decay rate. Although some calculations show that there 
exist large $c\overline{c}$ resonance contributions, which interfere with short
distance contributions \cite{long}, there is still a window in the invariant 
mass spectrum of the $e^+e^- $ pair for short distance contributions to be
dominant \cite{am}. Furthermore, window also exists for short 
distance to be dominant in the one lepton energy spectrum \cite{od}.
The QCD corrected coefficients of effective operators from $b \to se^+e^-$
are also important for the exclusive processes such as $B\to K(K^*) e^+e^-$.

The decay of $b\to se^+e^-$ and its large leading log QCD corrections 
have already been calculated in many papers \cite{Wis,Hou,Grin,Grig,cel}.
And also some efforts are made to give a next to leading log calculations 
\cite{Mis,AJB} which is estimated within 20\% contribution. All these efforts 
make it easy for experiments to detect this channel. However, all these papers 
do not include the QCD running from $m_{top}$ to $M_W$. Since the top quark is 
found to be 2-times heavier than W gauge boson \cite{CDF}, it needs a detail 
calculation for the effect of the QCD running from $m_{top}$ to $M_W$. 
 Further more, unlike the $b\to s \gamma$ case, here the electromagnetic
penguin is numerically less important than the box diagrams and Z boson 
penguin, although it is much enhanced by leading-log QCD corrections. 
Meanwhile, the important box diagrams and Z boson penguins have no 
leading-log QCD corrections as a running from $M_W$ to $m_b$ \cite{bur}. The only
leading-log QCD corrections may come from running from $m_{top}$ to $M_W$.

In the present paper, by using effective field theory formalism in standard
model, we recalculate the $b \to s e^+e^-$ decay 
including QCD running from $m_{top}$ to $M_W$, in addition to corrections
from $M_W$ to $m_b$, so as to give a complete leading log results. 
First in the next section, we integrate out the top quark, generating a 
five-quark effective hamiltonian. By using the renormalization group equation, 
we run the effective field theory down to the W-scale, so as to give out the 
QCD corrections from $m_{top}$  to $M_W$. In section 3, the weak gauge bosons 
are integrated out at  $M_W$ scale. Then we continue running the effective 
hamiltonian down to b-quark scale to include QCD corrections from $M_W$ to 
$m_b$. In section 4, the differential branching ratio is given as a function of
the invariant mass of the $e^+e^-$ pair. Our results will be useful for 
experiments to distinguish backgrounds like $c\bar c$ resonance. Section 5 is 
a short summary.

\section{QCD Corrections from $\mu =m_{top}$ to $\mu =M_W$ Scale}

At first, in the standard model Lagrangian, we integrate out the top quark, 
generating an effective theory, introducing dimension-5 and dimension-6 
effective operators as to include effects of the absent top quark.
Higher dimension operators are suppressed by a factor of $p^2/m_t^2$, where
$p^2$ characterize the interesting external momentum of b quark
$p^2\sim m_b^2$. For leading order of $m_b^2/m_t^2$, dimension-6
operators are good enough to make a complete basis of operators:
\begin{eqnarray}
O_{LR}^1  & =  &  -\frac{1}{16\pi^2} m_b \overline{s}_L D^2 b_R,
\nonumber\\
O_{LR}^2  &  =  &  \mu^{\epsilon/2} \frac{g_3}{16\pi^2} 
	m_b \overline{s}_L \sigma^{\mu\nu} X^a b_R G_{\mu\nu}^a,
\nonumber\\
O_{LR}^3  &  =  &  \mu^{\epsilon/2} \frac{e }{16\pi^2} 
	m_b \overline{s}_L \sigma^{\mu\nu} b_R	F_{\mu\nu},
\nonumber\\
Q_{LR}  &  =  &  \mu^{\epsilon} g_3^2 m_b 
	\phi_{+}\phi_{-} \overline{s}_L b_R,
\nonumber\\
P_L^{1,A}  &  =  &  -\frac{i}{16\pi^2}  \overline{s}_L 
  T_{\mu\nu\sigma}^A D^{\mu} D^{\nu} D^{\sigma} b_L,\nonumber\\
P_L^2 & = & \mu^{\epsilon/2} \frac{e Q_b}{16\pi^2}  \overline{s}_L 
	\gamma^{\mu} b_L \partial^{\nu} F_{\mu\nu},\nonumber\\
P_L^4  &  =  &  i \mu^{\epsilon/2} \frac{e Q_b}{16\pi^2}  
	\tilde{F}_{\mu\nu} 
	\overline{s}_L \gamma^{\mu} \gamma^5 D^{\nu} b_L,\nonumber\\
R_L^1 &  =  &  i \mu^{\epsilon} g_3^2 \phi_{+}\phi_{-} \overline{s}_L 
	\not \!\! D b_L,\nonumber\\
R_L^2  &  =  &  i \mu^{\epsilon} g_3^2(D^{\sigma} \phi_+) \phi_{-} 
	\overline{s}_L\gamma_{\sigma} b_L,
\nonumber\\
W_{LR} &  =  & -i \mu^{\epsilon} g_3^2 m_b W^{\nu}_{+}W_{-}^{\mu} 
	\overline{s}_L \sigma_{\mu \nu} b_R, \nonumber\\ 
W_L^1  &  =  &  i \mu^{\epsilon} g_3^2 W^{\nu}_{+}W_{-}^{\mu} 
	\overline{s}_L 
	\gamma _{\mu}\not \!\! D \gamma _{\nu} b_L,\nonumber\\
W_L^2  &  =  &  i \mu^{\epsilon} g_3^2(D^{\sigma} W^{\nu}_+) W^{\mu}_{-} 
	\overline{s}_L \gamma_{\mu} \gamma_{\sigma} \gamma_{\nu} b_L,
\nonumber\\
W_L^3  &  =  &  i \mu^{\epsilon} g_3^2 W_{+\mu} W^{\mu}_{-} 
\overline{s}_L \stackrel{\leftrightarrow}{\not \!\! D} b_L, \nonumber\\
W_L^4  &  =  &  i \mu^{\epsilon} g_3^2 W^{\nu}_+ W^{\mu}_{-} 
\overline{s}_L (\stackrel{\leftrightarrow}{D}_{\mu}\! \gamma_{\nu} +
	 \gamma_{\mu}\! \stackrel{\leftrightarrow}{D}_{\nu} ) b_L.\nonumber\\
Z_L^1  &  =  &  i \mu^{\epsilon} g_3^2 \frac{e}{\cos \theta_w \sin \theta_w}
	Z^{\mu} \overline{s}_L 
	\gamma _{\mu} b_L\phi^+ \phi^-,\nonumber\\
Z_L^2  &  =  &  i \mu^{\epsilon} g_3^2 \frac{e}{\cos \theta_w \sin \theta_w}
	Z^{\nu} \overline{s}_L \gamma_{\mu} \gamma_{\nu} \gamma_{\sigma} b_L
	W^{\mu}_- W^{\sigma}_+,
\nonumber\\
Z_{L}^3  &  =  &  i \mu^{\epsilon} \frac{e}{16\pi^2 \cos \theta_w \sin 
	\theta_w} M_W^2 Z_{\nu}  
	\overline{s}_L \gamma^{\nu} b_L, \nonumber\\
O_9 &=& (e^2/16\pi^2) ( \overline{s}_L\gamma^{\mu} b_L ) \overline{e} 
\gamma_{\mu} e,\nonumber\\
O_{10} &=& (e^2/16\pi^2) ( \overline{s}_L\gamma^{\mu} b_L ) \overline{e} 
\gamma_{\mu} \gamma_5 e.\label{operator}
\end{eqnarray}
Here $\overline{s}_L\! \stackrel{\leftrightarrow}{D}_{\mu} 
\!\gamma_{\nu} b_L$
stands for $(\overline{s}_L D_{\mu} \gamma_{\nu} b_L +(D_{\mu}
 \overline{s}_L)
 \gamma_{\nu} b_L)$ and the covariant derivative is defined as
$$D_{\mu}=\partial_{\mu}-i\mu^{\epsilon/2}g_3 X^a G_{\mu}^{a} -
i \mu^{\epsilon/2}eQ A_{\mu},$$
with $g_3$ denoting the QCD coupling constant.
The tensor $T_{\mu\nu\sigma}^A$ appearing in $P_L^{1,A}$
 assumes the following
Lorentz structure, the index $A$ ranging from 1 to 4:
$$
\begin{array}{ll}
        T_{\mu\nu\sigma}^1=g_{\mu\nu} \gamma_{\sigma},
&T_{\mu\nu\sigma}^2=g_{\mu\sigma} \gamma_{\nu},\nonumber\\
	T_{\mu\nu\sigma}^3=g_{\nu\sigma} \gamma_{\mu},
&T_{\mu\nu\sigma}^4=-i \epsilon_{\mu\nu\sigma\tau} 
	\gamma^{\tau} \gamma_5.
\end{array}
$$
The subscript $L$ and $R$ in the above formula denote left-handed and 
right-handed quarks, respectively. Here operators involving $bsz\phi W$ are 
not included because their coefficients are suppressed by $m_b/m_t$ and they 
do not mix with other operators. Then we can write down our intermediate
effective hamiltonian:
\begin{equation}
{\cal H}_{eff}=2 \sqrt{2} G_F V_{tb}V_{ts}^*\displaystyle \sum _i
C_i(\mu)O_i(\mu), \label{eff}
\end{equation}
where $V_{ij}$ represents the $3 \times 3$ unitary Kobayashi-Maskawa matrix 
elements.

The coefficients $C_i(\mu =m_{top})$ can be derived through matching Green
functions calculated from the standard model with that from the intermediate
effective theory \cite{lcd}. Keeping only leading order of $p^2/m_t^2$, the 
coefficients relevant to $b\to s \gamma$ decay are already given
in ref.\cite{lcd}, here we only give the new ones:
\begin{eqnarray}
C_{Z_L^1} &=&  \frac{1}{g_3^2} \left(\frac{1}{2} - \frac{2}{3}\sin ^2 
\theta \right),\\
C_{Z_L^2} &=& -\frac{1}{g_3^2} \frac{2}{3} 
	\sin^2 \theta _w \delta,\\
C_{Z_L^3} &=& \frac{1-6\delta}{2\delta(1-\delta)}-2\delta
	+\frac{1+2\delta-20\delta^2+12\delta^3}{2(1-\delta)^2}
	\log \delta -2\log \delta \\
	&+&\sin^2\theta_w\left( \frac{1}{3}+2\delta-\frac{1}{3}
	\log\delta-\frac{14}{3}\delta\log\delta\right),\nonumber\\
C_{O_9} &=& -\frac{1}{2\sin^2 \theta _w} \left ( \frac{ \frac{1}{4} 
	\delta + \frac{1}{4} \delta^2 }{ 1-\delta} 
	+\frac{ \delta -\delta^2 +\frac{1}{2} \delta^3}{ (1-\delta)^2}
	\log \delta \right), \\
C_{O_{10}} &=& \frac{1}{2\sin^2 \theta _w} \left ( \frac{ \frac{1}{4} 
	\delta + \frac{1}{4} \delta^2 }{ 1-\delta} 
	+\frac{ \delta -\delta^2 +\frac{1}{2} \delta^3}{ (1-\delta)^2}
	\log \delta \right),
\end{eqnarray}
with $\delta = M_W^2/m_t^2$.
The renormalization group equation satisfied by 
the coefficient functions $C_i(\mu)$ is
\begin{equation}
\mu \frac{d}{d\mu} C_i(\mu)=\displaystyle\sum_{j}(\gamma^{\tau})_{
ij}C_j(\mu).\label{ren}
\end{equation}
 The anomalous dimension matrix $\gamma_{ij}$
is calculated in practice by requiring 
renormalization group equations for Green functions 
with insertions of composite operators 
to be satisfied order by order in perturbation theory.

Only the last five operators in equation (\ref{operator}) are different 
from that of $b\to s \gamma$ case \cite{lcd}. 
After evaluating the loop diagrams, we get the
leading order anomalous dimensions for each of the operators
in our basis. Since there are so many operators, it is a very large
matrix of anomalous dimensions \cite{lcd}. Here we only list the part 
new from that of $b\to s\gamma$ case:
\begin{equation}
\begin{array}{rccl}
  \gamma=
   & \begin{array}{c} \\  R_L^2\\ W_L^2\\ Z_L^1 \\Z_L^2 \\Z_L^3
\\O_9\\ O_{10}  \end{array}
   & \begin{array}{c} 
  	\begin{array}{ccccccccccccccccccc} 
  R_{L}^2 & W_L^{2} && Z_L^1 & Z_L^2 &&&& Z_L^3 &&&&&
O_9 &&& O_{10} &
  	\end{array} \\
  
	\left(\begin{array}{ccccccc} 
         \frac{23}{48\pi^2} & 0 & 0& 0 & -\frac{1}{2}+\sin^2\theta_w 
& 0 & 0\\
	  0 & \frac{23}{48\pi^2}  & 0 & 0 & -6+6\sin^2\theta_w& 
\frac{1}{4\sin^2 \theta_w } & \frac{-1}{4\sin^2 \theta_w } \\
	  0 & 0 &   \frac{23}{48\pi^2}  & 0 & 1 & 0 & 0\\
	  0 & 0 &  0   & \frac{23}{48\pi^2} & 2 & 0 & 0\\
	  0 & 0 &  0   & 0 & 0  & 0 & 0\\
	  0 & 0 &  0   & 0 & 0  & 0 & 0\\
	  0 & 0 &  0   & 0 & 0  & 0 & 0\\
\end{array}\right)

     \end{array}

   &  \displaystyle{ 2g_3^2 }.

\end{array}\label{weak}
\end{equation}

The solution to renormalization group equation (\ref{ren}) 
appears in obvious matrix notation as
\begin{equation}
C(\mu_2)=\left[\exp\int_{g_3(\mu_1)}^{g_3(\mu_2)}dg\frac
{\gamma^T(g)}{\beta(g)}\right] C(\mu_1).\label{solu}
\end{equation}
After inserting anomalous dimension matrix (\ref{weak}),
we can have the coefficients of operators at $\mu=M_W^+$,
where the W boson has not been integrated out. 
\begin{eqnarray}
C_{P_L^2}(M_W^+) &=& C_{P_L^2}(m_t) +\frac{81}{226} (\zeta ^{113/138}-1)
\left[ C_{P_L^{1,2}}(m_t)+  C_{P_L^{1,4}}(m_t)\right] \nonumber\\
&&-\frac{1}{2}
 g_3^2 (m_t) \left[C_{R_L^2} (m_t) +2C_{W_L^2} (m_t)\right] 
\log \delta ,\nonumber\\
C_{P_L^4}(M_W^+) &=& C_{P_L^4}(m_t) +12
 g_3^2 (m_t) C_{W_L^1} (m_t)\log \delta ,\nonumber\\
C_{Z_i}(M_W^+) &=& \zeta C_{Z_i}(m_t) ,~~~ i=1,2,\nonumber\\
C_{Z_3}(M_W^+) &=& C_{Z_3}(m_t) + g_3^2 (m_t) C_{Z_L^1} (m_t) \log \delta
+ 2 g_3^2 (m_t) C_{Z_L^2} (m_t)\log \delta \nonumber\\
&&- (\frac{1}{2} -\sin ^2\theta_w)g_3^2 (m_t)
C_{R_L^2}(m_t)\log \delta -6(1-\sin^2 \theta_w) g_3^2 (m_t)
C_{W_L^2} (m_t)\log \delta,\nonumber\\
C_{O_9}(M_W^+) &=& C_{O_9}(m_t) + \frac{1}{4
\sin^2 \theta_w} g_3^2 (m_t) C_{W_L^2} (m_t)\log \delta,\nonumber\\
C_{O_{10}}(M_W^+) &=& C_{O_{10}}(m_t) -\frac{1}{4
\sin^2 \theta_w} g_3^2 (m_t) C_{W_L^2} (m_t)\log \delta ,
\end{eqnarray}
with $\zeta = \alpha_s(m_t)/\alpha_s(M_W)$.
The coefficients of other operators at $\mu=M_W^+$ like 
 $P_L^{1,2}$, $P_L^{1,4}$, $W_L^2$, $R_L^2$, are given at the 
appendix of ref.\cite{lcd2}.

\section{QCD Corrections from $\mu=M_W$ to $\mu =m_b$ Scale}

In order to continue running the operator coefficients down to lower scales, 
one must integrate out the weak gauge bosons W, Z and would-be
Goldstone bosons $\phi$ at $\mu=M_W$ scale. The matching conditions involving
four-quark operators and photon, gluon penguin diagrams are the same
as $b\to s \gamma$ case\cite{lcd}. We here only display 
the following relations between coefficient functions just below(-) and 
above(+) $\mu=M_W$ scale, which is relevant to $b \to s e^+ e^-$:
\begin{eqnarray}
C_{P_L^{1,2}}(M_W^-)&=&C_{P_L^{1,2}}(M_W^+) - 7/9, \nonumber\\
C_{P_L^{1,4}}(M_W^-)&=&C_{P_L^{1,4}}(M_W^+) + 1, \nonumber\\
C_{P_L^2}(M_W^-)&=&C_{P_L^2}(M_W^+) 
	-g_3^2(M_W)C_{W_L^2}(M_W^+) - 3/2, \nonumber\\
C_{P_L^4}(M_W^-)&=&C_{P_L^4}(M_W^+) + 9,\nonumber\\
C_{O_9}(M_W^-)  &=&  C_{O_9}(M_W^+) -\frac{g_3^2}{8\sin^2\theta_w}
	C_{W_L^2}(M_W^+) -\frac{1}{4\sin^2\theta_w}+\frac{4\sin^2
	\theta_w-1}{4\sin^2\theta_w}C(M_W)-\frac{1}{3}D(M_W)-\frac{4}
	{9},\nonumber\\
C_{O_{10}}(M_W^-)  &=&  C_{O_{10}}(M_W^+) +\frac{g_3^2}{8\sin^2\theta
	_w}C_{W_L^2}(M_W^+) +\frac{1}{4\sin^2\theta_w}+\frac{1}
	{4\sin^2\theta_w}C(M_W).\label{c2}
\end{eqnarray}
The first two terms in the right side of $C_{O_9}(M_W^-)$ and 
$C_{O_{10}}(M_W^-)$ arise from box diagrams with W boson couples to top
quark; the third terms arise from box diagrams with W boson coupling to charm 
quark. The function $C(M_W)$ arise from graphs with a Z gauge boson coupling
to the $e^+e^-$ pair, while the $D(M_W)$ function arises from graphs
with a photon couples to the $e^+e^-$ pair. They are defined as,
\begin{eqnarray}
C(M_W)&=&C_{Z_L^3}(M_W^+)-g_3^2C_{Z_L^1 }(M_W^+)+(\frac{1}{2}-\sin^2
\theta_w)g_3^2C_{R_L^2}(M_W^+)+2\cos^2\theta_wg_3^2C_{W_L^2}(M_W^+),
\nonumber\\
D(M_W)&=&C_{P_L^2}(M_W^-)-\frac{1}{2}
C_{P_L^{1,2}}(M_W^-)-\frac{1}{2}C_{P_L^{1,4}}(M_W^-)+
\frac{1}{2}C_{P_L^{4}}(M_W^-).
\end{eqnarray}

In addition to these, there are new four-quark operators from
integrating out the W boson\cite{Grin,Mis}:
\begin{eqnarray}
O_1&=&(\overline{c}_{L\beta} \gamma^{\mu} b_{L\alpha})
	    (\overline{s}_{L\alpha} \gamma_{\mu} c_{L\beta}),
\nonumber\\
O_2&=&(\overline{c}_{L\alpha} \gamma^{\mu} b_{L\alpha})
	    (\overline{s}_{L\beta} \gamma_{\mu} c_{L\beta}),
\nonumber\\
O_3&=&(\overline{s}_{L\alpha} \gamma^{\mu} b_{L\alpha})
	    [(\overline{u}_{L\beta} \gamma_{\mu} u_{L\beta})+...+
	    (\overline{b}_{L\beta} \gamma_{\mu} b_{L\beta})],
\nonumber\\
O_4&=&(\overline{s}_{L\alpha} \gamma^{\mu} b_{L\beta})
	    [(\overline{u}_{L\beta} \gamma_{\mu} u_{L\alpha})+...+
	    (\overline{b}_{L\beta} \gamma_{\mu} b_{L\alpha})],
\nonumber\\
O_5&=&(\overline{s}_{L\alpha} \gamma^{\mu} b_{L\alpha})
	    [(\overline{u}_{R\beta} \gamma_{\mu} u_{R\beta})+...+
	    (\overline{b}_{R\beta} \gamma_{\mu} b_{R\beta})],
\nonumber\\
O_6&=&(\overline{s}_{L\alpha} \gamma^{\mu} b_{L\beta})
	    [(\overline{u}_{R\beta} \gamma_{\mu} u_{R\alpha})+...+
	    (\overline{b}_{R\beta} \gamma_{\mu} b_{R\alpha})],
\end{eqnarray}
with coefficients
$$ C_i(M_W)=0, \;\; i=1,3,4,5,6, \;\; C_2(M_W)=-1.$$

   There should also be two magnetic moment operators relevant to $b\to
s \gamma$, 
\begin{eqnarray}
O_{7}  &  =  &   \frac{e }{16\pi^2} 
	m_b \overline{s}_L \sigma^{\mu\nu} b_R	F_{\mu\nu},\\
O_{8}  &  =  &   \frac{g_3}{16\pi^2} 
	m_b \overline{s}_L \sigma^{\mu\nu} X^a b_R G_{\mu\nu}^a,
\end{eqnarray}
their coefficients at $\mu=M_W^-$ are given in ref.\cite{lcd}.

The operator basis now consists of 10 operators $O_1-O_{10}$.
The  effective hamiltonian appears just below the W-scale as
\begin{equation}
{\cal H}_{eff} =\frac{4G_F}{\sqrt{2}} V_{tb}V_{ts}^*  
	\displaystyle{\sum_{i=1}^{10} }C_i (M_W^-) O_i(M_W^-).
\end{equation}	 

If the QCD corrections from $m_{top}$ to $M_W$ are ignored [by 
setting $\alpha_s(m_t)=\alpha_s(M_W)$
 in eqn.(\ref{c2})], the above results(\ref{c2})
 would reduce to the previous results \cite{Grin,Mis} exactly,
where the top quark and W bosons are integrated out together. This is a
necessary consistent check.

The running of the coefficients of operators from $\mu=M_W$ to $\mu=m_b$
was well described in previous papers\cite{Grin,Mis}. After  running we have 
the coefficients of operators at $\mu=m_b$ scale \cite{ali}. 
\begin{eqnarray}
C_7(m_b) &= & \eta^{16/23}C_7(M_W) +\frac{8}{3} ( \eta^{14/23}
-\eta^{16/23} ) C_8(M_W) +C_2(M_W) \displaystyle \sum _{i=1}^{8} h_i 
\eta^{a_i},\\
C_9(m_b) &= & C_9(M_W)+\frac{C_2(M_W)}{\alpha_s(m_b)}\left(-0.0938/\eta 
+\sum _{i=1}^{8} p_i \eta ^{a_i }\right),\\
C_{10}(m_b) &= & C_{10}(M_W),
\end{eqnarray}
with $\eta = \alpha_s(M_W) /\alpha_s (m_b)$,
$ h_i$,$a_i $ and $p_i $ defined in ref.\cite{ali}.

\section{Results}

In a spectator model, the inclusive decay $\overline{B} \rightarrow X_s e^+e^-$ is mainly 
contributed from the b quark decay $b\rightarrow s e^+e^-$. To leading
order, the differential decay rate $d\Gamma(
\overline{B} \rightarrow X_s e^+e^-)/d \hat s$, where $\hat s =( p_e^+
+p_e^-)^2/m_b^2$, is given by:
\begin{eqnarray}\label{wid}
\frac{1}{\Gamma (\overline{B} \rightarrow X_c e \overline{\nu})}
 \frac{d}{d\hat s} \Gamma(\overline{B} \rightarrow X_s e^+e^-)
&=& \frac{\alpha_{QED}^2}{4\pi^2
 f (m_c/m_b)} (1-\hat s)^2 \left[ (1+2\hat s) 
 \left( |C_9^{eff}|^2 +C_{10}^2\right)\right .\nonumber\\
&+& \left .4\left( 1+\frac{2}{\hat s} \right) |C_7|^2 +12 C_7
 Re(C_9^{eff}) \right ].
\end{eqnarray}
Here  
\begin{eqnarray}
C_9^{eff}&=& C_9(m_b) +g(m_c/m_b, \hat s) (3 C_1+C_2+3C_3+C_4+3C_5+C_6)\\
&& -\frac{1}{2} g( 1, \hat s)( 4C_3+4C_4+3C_5+C_6) -\frac{1}{2} 
g( 0, \hat s)( C_3+3C_4),
\end{eqnarray}
where $g(m_c/m_b,\hat{s})$ arises from the one-loop matrix element of the four-
quark operators, which can be written as \cite{Grin}
\begin{equation}
g(z,\hat s) = -\frac{8}{9}\ln z +\frac{8}{27}+\frac{4}{9}\beta
 -\frac{2}{9} \left(2+\beta\right) \sqrt{|1-\beta|} \left\{
	\begin{array}{ll}
	\ln \left| \frac{\sqrt{1-\beta} +1}{\sqrt{1-\beta }-1}\right|
	+i\pi, & \beta <1, \\
	2 \arctan ( 1/\sqrt{\beta-1} ), & \beta >1,
	\end{array}  \right.
\end{equation}
with $\beta=4z^2/\hat s$.
The factor of $f(m_c/m_b)$ arises because of the dependence of the semileptonic
decay rate on the nonnegligible ratio $m_c/m_b$, and is given by
$$f(x)=1-8x^2+8x^6-x^8-24x^4\ln x.$$ In eqn.(\ref{wid}) the masses of the 
electron and the strange quark were neglected.

The semileptonic decay $\overline{B} \to X_c e\overline{\nu}$ \cite{Cabi} is 
used to eliminate large uncertainties of $m_b^5$ in the decay width formula.  
The dependence on the weak mixing angles also cancels out.
If we take experimental result $Br(\overline{B} \to
X_c e\overline{\nu} ) =10.8\% $ \cite{data}, the branching ratios of
$\overline{B} \to X_s e^+e^-$ is found.

Taking values as $M_W=80.22$GeV, $m_t=175$GeV, $m_b=4.8$GeV 
and the QCD coupling constant $\alpha_s(m_Z)=0.117$ \cite{data},
the differential branching ratios are depicted in Fig.1  as a function of
$\hat s$. It is shown that the QCD corrections from $m_t$ to $M_W$ 
slightly suppress the $b\to s 
e^+e^-$ differential decay rate for a large
invariant mass of the $e^+e^- $ pair; while 
enhance it for a very small invariant mass of the $e^+e^- $ pair.
Since the enhancement is larger than the suppression in the invariant
 mass spectrum of the $e^+e^- $ pair, the total branching ratio is found 
to be slightly enhanced about 4\% comparing with the one without QCD correction from  $m_{top}$ to $M_W$.

\section{Conclusion}

As a conclusion, we have given the full leading log QCD 
corrections (including
QCD running from $m_{top}$ to $M_W$) to $b\to s e^+e^-$ decay in the 
standard model.

The QCD correction from $m_t$ to $M_W$ slightly suppresses the $b\to s 
e^+e^-$ differential decay rate for a large
invariant mass of the $e^+e^- $ pair. While the integrated width is slightly
enhanced comparing with that  without the 
QCD running from  $m_{top}$ to $M_W$.

Although this result is not quite different from
the previous calculations, our improvement lies in reducing some 
theoretical uncertainties. 

\section*{Acknowledgement}

C.D. L\"u thanks X.Y. Li for 
helpful discussions.

\section*{Figure Captions}
\parindent=0pt

Fig.1 Differential branching ratios of $b\to s e^+e^-$, as a function
of $s=(p^++p^-)^2/m_b^2$. The solid line denotes the result with full 
QCD corrections, while the dashed one corresponds to result without QCD
corrections from $m_t$ to $M_W$.

\newpage
\begin{figure}
\centerline{\epsffile{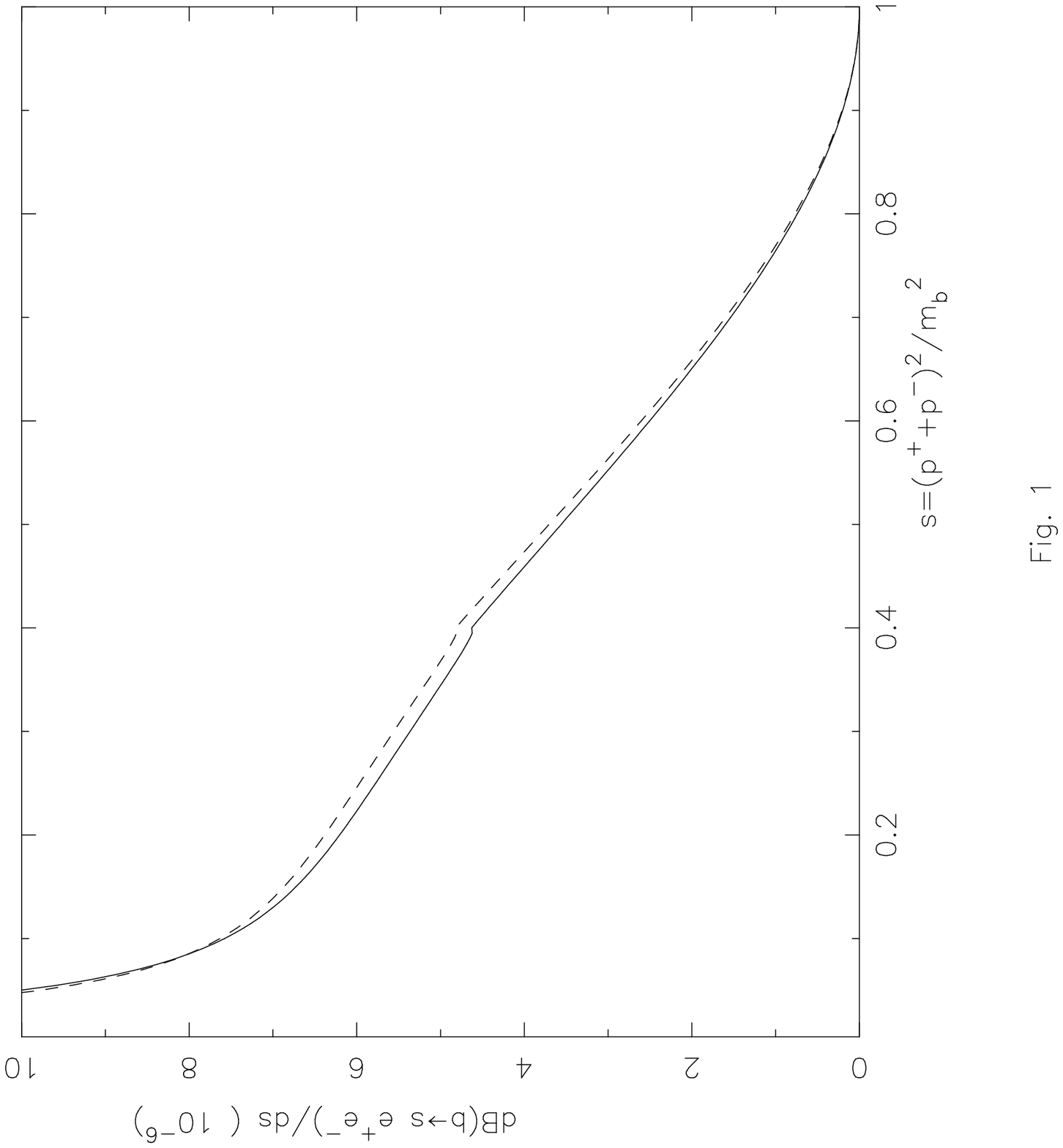}}
\end{figure}

\end{document}